%% file: ms.tex
\RequirePackage{fix-cm}
\documentclass[twocolumn,epjc3]{svjour3}
\RequirePackage[T1]{fontenc}

\smartqed  

\RequirePackage{graphicx}
\RequirePackage{mathptmx}      
\RequirePackage{flushend}
\RequirePackage[numbers,sort&compress]{natbib}
\RequirePackage[colorlinks,citecolor=blue,urlcolor=blue,linkcolor=blue]{hyperref}
\journalname{Eur. Phys. J. C}

\graphicspath{{./graphics/}}

\usepackage{isotope}
\usepackage{dcolumn}
\usepackage{bm}
\usepackage{epsfig}
\usepackage{pdflscape}
\usepackage{subfigure} 
\usepackage{multirow}
\usepackage{rotating}
\usepackage{tabu}
\usepackage{longtable}
\usepackage{booktabs}
\usepackage{textcomp}
\usepackage{microtype}

\usepackage{siunitx}
\DeclareSIUnit{\year}{a}
\DeclareSIUnit{\photoelectron}{PE}
\DeclareSIUnit{\ppb}{ppb}
\DeclareSIUnit{\ppt}{ppt}
\DeclareSIUnit{\ppq}{ppq}
\DeclareSIUnit{\ccm}{\cubic\cm}
\DeclareSIUnit{\mbar}{\milli\bar}
\DeclareSIUnit{\dru}{DRU}

\usepackage{ulem}

\newcommand*{\perc}[1]{\SI{#1}{\percent}}

\input{affiliations}

\begin{document}
\input{ib_authorlist_epjc}

\title{Intrinsic backgrounds from Rn and Kr in the XENON100 experiment}
\date{Received: date / Revised version: date}

\maketitle
\begin{abstract}
In this paper, we describe the XENON100 data analyses used to assess the
target-intrinsic background sources radon (\isotope[222]{Rn}), thoron
(\isotope[220]{Rn}) and krypton (\isotope[85]{Kr}). We detail the event
selections of high-energy alpha particles and decay-specific delayed
coincidences. We derive distributions of the individual radionuclides
inside the detector and quantify their abundances during the main
three science runs of the experiment over a period of \SI{\sim 4}{years}, from January 2010 to January 2014. We compare our results to external
measurements of radon emanation and krypton concentrations where we find good agreement. We report an observed reduction in concentrations of radon daughters that we attribute to the plating-out of charged ions on the negatively biased cathode. 
%
\end{abstract} 
%

\section{Introduction}
\label{sec:intro}
Liquid noble gas detectors play an important role in rare-event search experiments
looking for dark matter interactions or neutrinoless double beta decay~\cite{Aprile:2009dv}. 
One of their key features is ease of scalability. With larger target masses,
external radioactivity can be better shielded through fiducialization. This is not the case for internal backgrounds that are intrinsic to the liquid gas target. First, these are medium- to long-lived radioisotopes of
the target itself. For instance, liquid argon detectors need to take special care to avoid
\isotope[39]{Ar}~\cite{Agnes:2015ftt}. In the case of liquid xenon detectors,
the two-neutrino double-beta emitter \isotope[136]{Xe} becomes relevant
at the multi-ton scale~\cite{Aprile:2015uzo}. Second, and more relevant for xenon detectors,
are the radionuclides from radon and krypton. Both elements are inert gases
that cannot be removed by established purification techniques based on hot gas
purifiers commonly used in the field~\cite{Dobi:2011dd,Aprile:2011dd,Akerib:2012ys,Cao:2014jsa}. Radon and
krypton dissolve in the liquid xenon target and cannot be excluded by standard
fiducialization techniques which otherwise allow the rejection of background~\cite{Aprile:2011dd}.
In this work we describe how krypton and radon backgrounds are assessed in
the XENON100 experiment. Dark matter data from the three main XENON100
science runs (SRs), with exposure times of 101, 223 and 153 days each, are examined. The runs themselves and the corresponding
detector conditions are outlined in Table~1 of~\cite{Aprile:2016swn}. 

\section{The XENON100 detector}
\label{sec:detector}
The XENON100 detector~\cite{Aprile:2011dd} is a cylindrical dual-phase time projection chamber (TPC) of \SI{30.5}{\centi\meter} height
and \SI{30.6}{\centi\meter} diameter. It is located at the Laboratori Nazionali del Gran Sasso (LNGS) and uses about \SI{62}{\kilo\gram} of
liquid xenon (LXe) as a target which is monitored by two arrays of photomultiplier tubes (PMTs), with one being at the top and one at the
bottom of the TPC. Its primary goal is to search for dark matter in the form of weakly interacting massive particles (WIMPs).

Incoming particles are detected via their interactions with the LXe, generating xenon scintillation
photons (S1 signal) as well as ionization electrons. The electrons then drift towards the top of the TPC
due to a homogeneous drift field applied across the LXe volume. At the top of the TPC, the electrons are accelerated into
a region of gaseous xenon (GXe) by an extraction field. Due to the moving electrons interacting with the GXe, proportional scintillation
photons are created (S2 signal)~\cite{dolgoshein:1970}. Both S1 and S2 signals, measured in photoelectrons (PE), are detected by the PMT arrays. The delay between the S1 and
the S2 signals of an interaction, combined with the hit pattern of the S2 signal on the top array, allows the reconstruction of all 3
coordinates of the interaction vertex. The $X$ and $Y$ coordinates are defined
relative to the TPC's central axis (where $X = Y = 0$) and are determined with a resolution of $\sigma_{X/Y} < \SI{3}{\milli\meter}$. The $Z$ coordinate has a resolution of $\sigma_{Z} < \SI{0.3}{\milli\meter}$ and is defined with respect to the liquid gas interface at
the top ($Z = 0$) and the cathode electrode, which is used to create the drift field, at the bottom of the TPC
($Z = \SI{-30.5}{\centi\meter}$).

In addition, the S2/S1 ratio allows discrimination between
nuclear recoils (NRs), which WIMPs are expected to induce, and electronic
recoils (ERs), produced by $\gamma$-rays and $\beta$-particles. For example,
in XENON100 WIMP analyses, \SI{99.75}{\percent} of ERs can be rejected at the price of an
energy-dependent NR acceptance of \SIrange{30}{50}{\percent} by utilizing
this feature~\cite{Aprile:2012vw}. As tails of the ER distribution contaminate
the NR region, it is of paramount importance to measure the abundance of
\isotope[222]{Rn} and \isotope[85]{Kr} and to estimate their impact on
the detector's sensitivity to WIMP interactions.

\section{Radon and thoron}
\label{sec:rn}

\begin{figure}
	\centering
	\includegraphics[keepaspectratio=true, width=0.45\textwidth]{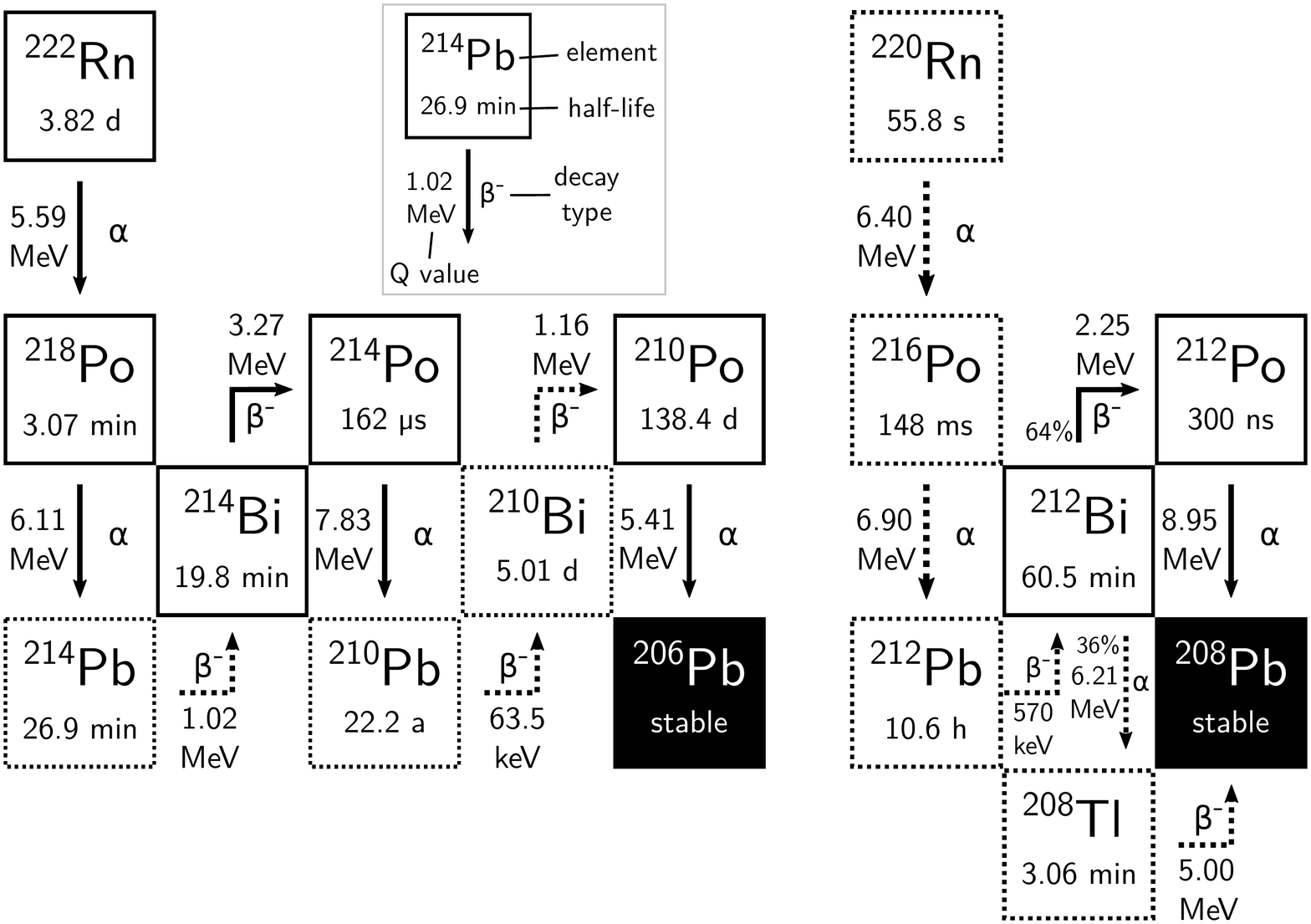}
	\caption{Illustrations of the \isotope[222]{Rn} (radon) and \isotope[220]{Rn} (thoron) chains (ignoring decay modes with a branching
	ratio $\leq \perc{0.1}$). Half-lives, branching ratios and $\mathcal{Q}$-values are taken from
	\protect{~\cite{Be:tab}}. Solid boxes mark the isotopes that are quantified in
    this work}
	\label{fig:rn_chains}
\end{figure}

The decays of \isotope[222]{Rn} (radon)
and \isotope[220]{Rn} (thoron), as well as their daughters,
are illustrated in Figure~\ref{fig:rn_chains} with the
half-lives, $\mathcal{Q}$-values and branching ratios used
throughout this work \cite{Be:tab}. Radon and thoron are produced in the decay chains of the primordial
nuclides \isotope[238]{U} and \isotope[232]{Th}, respectively. Both of these nuclides are present,
at least at trace level, in all materials, making it necessary to carefully screen and select all detector
components~\cite{Aprile:2011ru}. The radon concentration of the air underground at LNGS,
emanating from the surrounding rock, has been found
to be of $\mathcal{O}(\SI{100}{\becquerel\per\meter\cubed})$~\cite{Aprile:2012vw}. For this reason, the inner cavity of the detector's
shield is continuously flushed with boil-off nitrogen~\cite{Aprile:2011dd},
minimizing the amount of ambient radon and thoron that could potentially enter.

Levels of radon and thoron inside the LXe target of XENON100 are
determined by the emanation of either isotope from surfaces inside the
detector and the xenon purification system. Additionally, one period before SR1 and two periods during SR2 and SR3,
were identified where air leaks of $\mathcal{O}(\SI{e-3}{\milli\bar\litre\per\second})$ and $\mathcal{O}(\SI{e-5}{\milli\bar\litre\per\second})$, respectively, developed at the purification system's diaphragm
pump, leading to a variation of the radon background
over time (see \cite{Aprile:2017yea} and Sec.~\ref{subsec:rn_results}).

After entering the LXe, radon and thoron are able to reach the fiducial
volume used for the WIMP search via diffusion and convection~\cite{Aprile:2016pmc}. As a
consequence, $\beta$-decaying daughter nuclides of both
isotopes can contribute to the
low-energy ER background. Contributions from $\alpha$-decays are not relevant because the
involved $\alpha$-particle energies are two orders of magnitude larger than the energies
expected from WIMP-induced NRs, which are at $\mathcal{O}(\SI{10}{\kilo\electronvolt})$~\cite{Aprile:2012vw}.

Some of the progeny of the chains' nuclei have short
half-lives compared to the event window of
XENON100, which has a length of \SI{400}{\micro\second} and is
centered on the triggering signal~\cite{Aprile:2011dd}. This aspect results in two
decays being recorded within the same event (delayed coincidence
signature). An example for this are decays of \isotope[214]{Bi} (radon chain) and
\isotope[212]{Bi} (thoron chain) which are followed by the decays of their polonium
daughters (BiPo coincidence). This causes multiple S1 and S2 signals to be present in an event,
making it possible to identify and reject them (see Sec.~\ref{subsec:bipo_event_selection}).

Of the $\beta$-decaying nuclides in either chain, many have a
significant likelihood of decaying under prompt emission of
$\gamma$-rays, which gives the same kind of signature as mentioned above. However, certain
$\beta$-decaying nuclides from either chain are able to  decay
without $\gamma$-ray emission. If they are well-separated in time
from accompanying decays, they are very likely to elude
identification. Such nuclides are \isotope[214]{Pb}, \isotope[210]{Pb} (radon chain) and
\isotope[212]{Pb} (thoron chain).

Radon and thoron concentrations in the LXe target can be inferred by
selecting and counting events from decays of their chains.
Especially suited for this task are $\alpha$-decays, as they
produce large, monoenergetic signals resulting in a distinct event
signature. Another explicit signature is the BiPo delayed coincidence
as outlined above. This coincidence has already been successfully utilized by,
for instance, the Borexino and SuperNEMO collaborations for estimating radioactive
background levels inside their detectors~\cite{Alimonti:1998aa,Gomez:2013dka} and by the XENON collaboration
for assessing the suitability of a thoron source for calibrating tonne-scale LXe detectors~\cite{Aprile:2016pmc}. The focus of this section is to describe the selection of
$\alpha$-decays and BiPo events. Results thereof are presented in
Sec.~\ref{subsec:rn_results}.

\subsection{Alpha event selection}
\label{subsec:alpha_event_selection}
For the analysis of $\alpha$-decaying nuclides in XENON100,
\isotope[222]{Rn} and \isotope[218]{Po} are used because they are the cleanest $\alpha$ populations available as explained further below.
\isotope[214]{Po} is covered in Sec.~\ref{subsec:bipo_event_selection} as part of the BiPo coincidence.

To select $\alpha$-decays, a set of cuts, based on criteria described in~\cite{Aprile:2012vw,Weber:2013tci}, is applied to the
data. We require at least one S1 signal with a minimum of two PMTs in coincidence. Any secondary S1 signal has to be below \SI{1600}{\photoelectron}
(motivated in~\cite{Weber:2013tci}) to avoid decay pileup and multi-scatters. In addition, at least one S2 signal must be present with
at least \SI{25}{\percent} of its area observed by the top PMT array to reject mis-identified signals. Due to the large signal sizes of $\alpha$-decays, which are found to be of $\mathcal{O}(\SI{e4}{\photoelectron})$ for S1s and $\mathcal{O}(\SI{e5}{\photoelectron})$ for S2s, acceptance losses due to the above mentioned cuts are considered to be negligible.

Finally, the detector volume used for this analysis is
restricted to $R = \sqrt{X^2+Y^2} < \SI{135}{\milli\meter}$ and $\SI{10}{\milli\meter} < Z < \SI{260}{\milli\meter}$ ($\hat{=}\ \SI{40.5}{\kilo\gram}$ LXe or \SI{65.3}{\percent} of the active volume).
This excludes regions close to the TPC walls, which suffer from reduced light and charge collection efficiencies, and regions with insufficient separation between \isotope[222]{Rn} and \isotope[218]{Po}.

\begin{figure}
	\centering
	\includegraphics[keepaspectratio=true, width=0.47\textwidth]{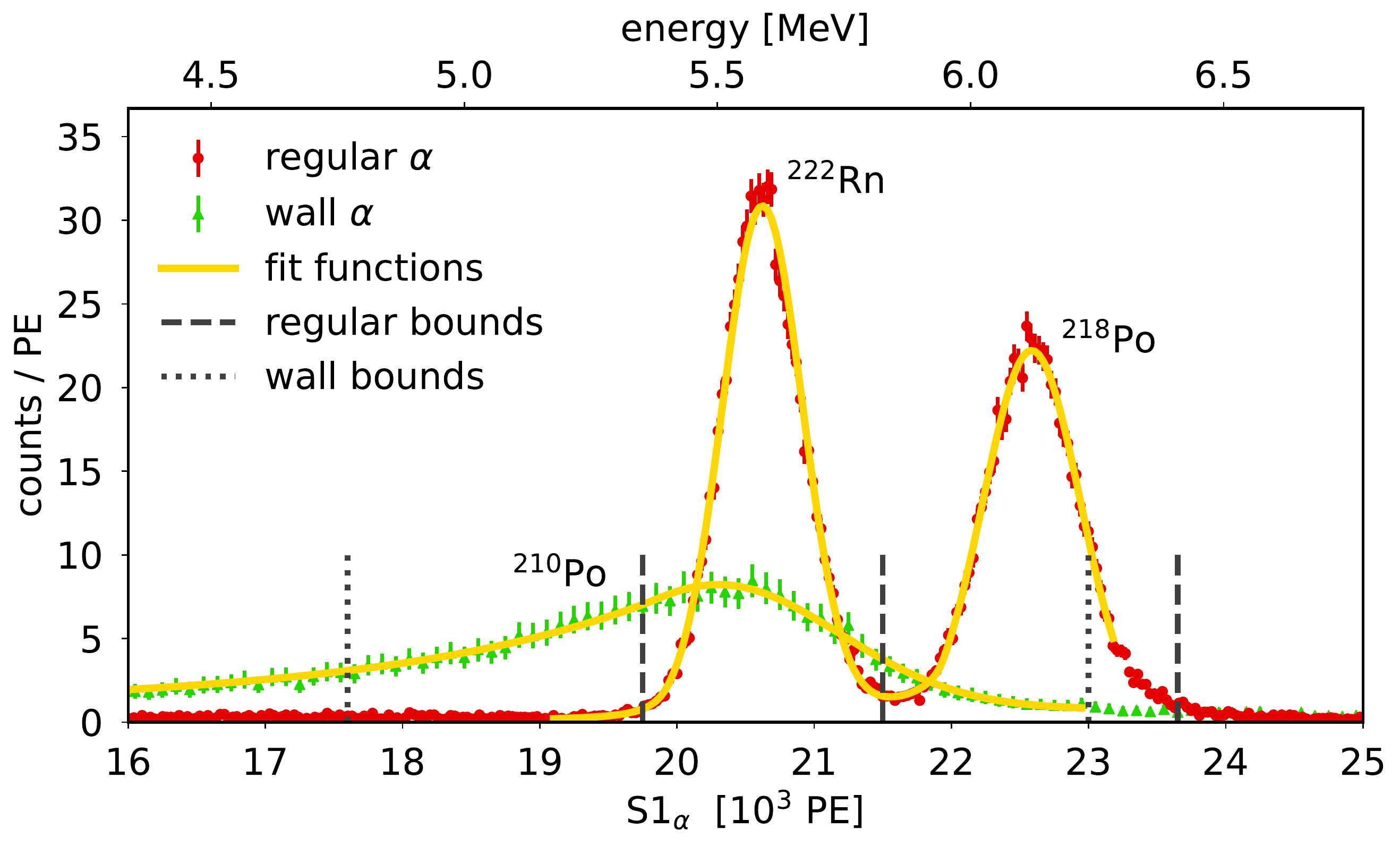}
	\caption{Corrected S1$_{\alpha}$ spectra (SR1). The left peak of the regular $\alpha$ selection spectrum
	(red, circular markers) is attributed to \isotope[222]{Rn}, the right one to \isotope[218]{Po}, with their bounds
	($3\sigma$ intervals) indicated by dashed lines. The tailing peak in the wall selection spectrum (green, triangular markers) is
	considered to belong to \isotope[210]{Po} and has its bounds indicated by dotted lines}
	\label{fig:s1_spectrum}
\end{figure}

A potential \isotope[210]{Po} population close to the PTFE wall enclosing the TPC is also selected for further studies.
Selection criteria are the same as above, with the following differences: $R \geq \SI{135}{\milli\meter}$
is required, and the largest S2 signal must be smaller than \SI{8e4}{\photoelectron}. The latter criterion is motivated by the observation of S2 signals well below those seen from \isotope[222]{Rn} and \isotope[218]{Po} for wall population events. Reduced
S2 signals correspond to charge losses which can result from, for example, decays happening close to or within the walls. In the latter case, decay
products can still enter the TPC, but lose energy in the process as they need to traverse the wall material.
The S1 signal cannot be used as the only parameter for nuclide discrimination in this case, as the detector's energy resolution is insufficient to
separate the peaks of \isotope[222]{Rn} and \isotope[210]{Po} in the S1 spectrum (see Figure~\ref{fig:s1_spectrum}).

The largest S1 and S2 signals are interpreted as belonging to an $\alpha$-decay and are correspondingly named
S1$_{\alpha}$ and S2$_{\alpha}$. A position-dependent area
correction for $\alpha$-decays~\cite{Weber:2013tci} is applied to S1$_{\alpha}$ in order to
account for PMT saturation, which affects both the observed signal size and position reconstruction. Looking at
the corrected S1$_{\alpha}$ spectrum for events happening at $R < \SI{135}{\milli\meter}$
(Figure~\ref{fig:s1_spectrum}, red circular markers), we identify two peaks, which are attributed to the $\alpha$-decays of
\isotope[222]{Rn} and \isotope[218]{Po}, respectively.

To determine peak positions and extents while accounting for slight tailing, we fit a sum of two Crystal Ball functions (defined in equation (F-1)
of~\cite{Gaiser:1982yw}) and a constant to the peaks. Events are classified as containing a \isotope[222]{Rn} or
\isotope[218]{Po} decay if the area of their S1$_{\alpha}$ is within $3\sigma$ of the respective peak mean (bounds as shown in
Figure~\ref{fig:s1_spectrum}). This choice is valid as the peaks are, in good approximation, symmetric. Fits are done separately for each SR
due to changes of detector parameters affecting positions and widths of both peaks~\cite{Aprile:2016swn}. In all SRs, the peak bounds determined according to this method do either not overlap, or overlap negligibly. In the latter case, the bound separating both peaks is determined by the arithmetic mean of the overlapping bounds in order to ensure events to be attributed to a single peak only. Leakage of the peaks beyond the boundaries assigned to them are estimated to be $< \SI{1}{\percent}$ and are thus considered negligible. For consistency, the same
procedure is utilized for the wall population, as it also shows a peak in the S1 spectrum (Figure~\ref{fig:s1_spectrum}, green triangular markers), using a single Crystal Ball function plus a constant for fitting.

In the thoron chain, the number of $\alpha$-decays from \isotope[220]{Rn} and \isotope[216]{Po} can,
in principle, be inferred from the S1$_{\alpha}$ spectrum via peak fitting (see~\cite{Aprile:2017kop}). However, in the XENON100 background, the
\isotope[218]{Po} peak overlaps the \isotope[220]{Rn} peak region due to insufficient energy resolution. In addition, there
is no indication of \isotope[216]{Po} being present in the S1$_{\alpha}$ spectrum of the fiducial volume used, while, at the same time, it is negligible
in the rest of the sensitive volume compared to the wall population. As thus no direct evidence of them exists in the fiducial volume, \isotope[220]{Rn} and \isotope[216]{Po} are not taken
into account in this analysis, even though they are present in the detector as demonstrated by \isotope[212]{Po} being measured, which belongs to the nuclei discussed in detail in
the following section.

\subsection{BiPo event selection}
\label{subsec:bipo_event_selection}
The decays of \isotope[214]{Bi} and \isotope[212]{Bi} are often recorded within the same event as the decays
of their daughter nuclei, \isotope[214]{Po} and \isotope[212]{Po}. This is due to the short half-life of the
polonium isotopes compared to the event window recorded by the XENON100 data acquisition system. The S1 signals generated by the $\beta$
decays of the bismuth isotopes (S1$_{\beta}$) are smaller than those generated by the $\alpha$-decays of the
polonium daughters (S1$_{\alpha}$), because the
$\beta$-decay $\mathcal{Q}$-values (see Figure~\ref{fig:rn_chains}) and ionization densities are lower than
those of the $\alpha$-decays~\cite{Aprile:2006kx}.

The result is a delayed coincidence signature of one S1 signal being followed by a larger one. For selecting such events, we require at least two S1 signals
with at least twofold PMT coincidence and the correct time order (S1$_{\beta}$ before S1$_{\alpha}$).
Both signals need to be larger than \SI{200}{\photoelectron} and S1$_{\alpha}$ has to pass a data quality cut on the fraction
of its area observed by the top PMT array to reject signals seen almost exclusively by the bottom array. Such a signal topology is virtually
impossible to occur for an $\alpha$-decay happening inside the TPC due to the large amounts of scintillation photons generated
(see Sec.~\ref{subsec:alpha_event_selection}).

In order to distinguish delayed coincidences from Bi and Po decays (called BiPos in the following) from either chain additional constraints are applied exploiting the fact that \isotope[212]{Po} has a
much shorter half-life than \isotope[214]{Po} ($T_{1/2} = \SI{300}{\nano\second}$
vs. $T_{1/2} = \SI{162}{\micro\second}$). \isotope[214]{BiPos} are selected by requiring S1$_{\alpha}$ to occur
at least \SI{7}{\micro\second} after S1$_{\beta}$, which removes more than \SI{99.99}{\percent} of
\isotope[212]{BiPo} events. For \isotope[212]{BiPos}, the time difference has to be between
\SI{0.5}{\micro\second} and \SI{2}{\micro\second}. The lower bound ensures that both signals are individually identified
with $\sim \SI{100}{\percent}$ efficiency by the data processor, while the upper bound removes about  \SI{99}{\percent} of \isotope[214]{BiPo} events.

Due to the possibility of $\gamma$-radiation accompanying the Bi-decays, the S2$_{\alpha}$ signal falling outside the
event window, and signal losses because of the spatial distribution of events as detailed in Sec.~\ref{subsec:rn_results}, no constraints are required on the number of S2 signals and
their parameters. In fact, as the number of S2 signals is expected to vary and event reconstruction is
not optimized for pairing S1 and S2 signals when multiple physical interactions are present, signal
matching has to be done separately. A match requires the absolute time difference between a pair of S2 signals to be within $\sim \SI{1}{\micro\second}$ of the one between
the S1 signals (detailed in~\cite{Cichon:2015}). The S2 which occurs earlier is assigned to S1$_{\beta}$. If no match is found,
the largest S2 is assigned to S1$_{\beta}$. We then recalculate positions and signal corrections (for S1$_{\alpha}$ and those
mentioned in~\cite{Aprile:2011dd}) for each event, as both depend on pairing
S1 with S2 signals. Events without any S2 signal are not rejected, but are assumed to have occurred in a charge-insensitive region such as below the cathode,
with a set of default coordinates assigned to them ($R = \SI{0}{\centi\meter}$, $Z = \SI{-30.5}{\centi\meter}$).

The data processor does not search for S1 signals occurring after a sufficiently large S2 signal
within the same event~\cite{Aprile:2011dd}. This behavior is intentional as the processor has been developed for the analysis of single
interaction events. However, this reduces the acceptance of the BiPo event selection because the S1$_{\alpha}$ signal
might occur after the first S2 signal of the Bi-decay which happens after S1$_{\beta}$ within the maximum drift time of
\SI{176}{\micro\second}~\cite{Aprile:2011dd}. This loss in acceptance as well as the one resulting from the finite size of the
event window is accounted for by summing up weights $\epsilon$ for each BiPo event, defined by
\begin{equation}
	\epsilon^{-1} = \exp\left(-\lambda\ \Delta t_{\textnormal{min}}\right)-\exp\left(-\lambda\ \Delta t\right).
	\label{eq:rn_eps}
\end{equation}
$\lambda$ is the decay constant of the corresponding polonium isotope, $\Delta t$ is the time difference between
S1$_{\beta}$ and the first S2 peak (or the end of the event window, if no S2 is present), and
$\Delta t_{\textnormal{min}}$ is the minimum time difference between S1$_{\alpha}$ and
S1$_{\beta}$ allowed by the selection criteria. Thus, the right side of equation
(\ref{eq:rn_eps}) is the probability of a polonium decay to occur within the given constraints in time.

Acceptance losses caused by the S1$_{\beta}$ size criterion, however, cannot be reliably predicted without depending on
measured \isotope[222]{Rn} and \isotope[220]{Rn} rates. This is caused by events that happen on the cathode, whose
S1 signals are shadowed by the cathode grid which results in a modification of the expected S1$_{\beta}$
spectrum. The relevance of cathode events is explained in the following section. Losses induced by other quality cuts are negligible.

\subsection{Results and discussion}
\label{subsec:rn_results}
\begin{figure*}
	\centering
	\includegraphics[keepaspectratio=true, width=\textwidth]{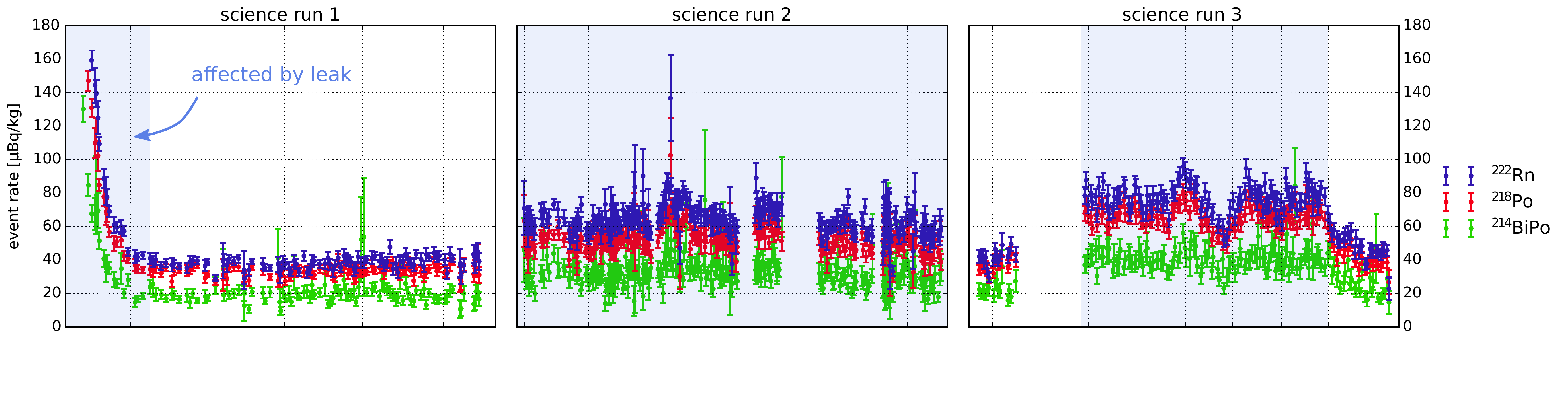}\\[-25pt]
	\includegraphics[keepaspectratio=true, width=\textwidth]{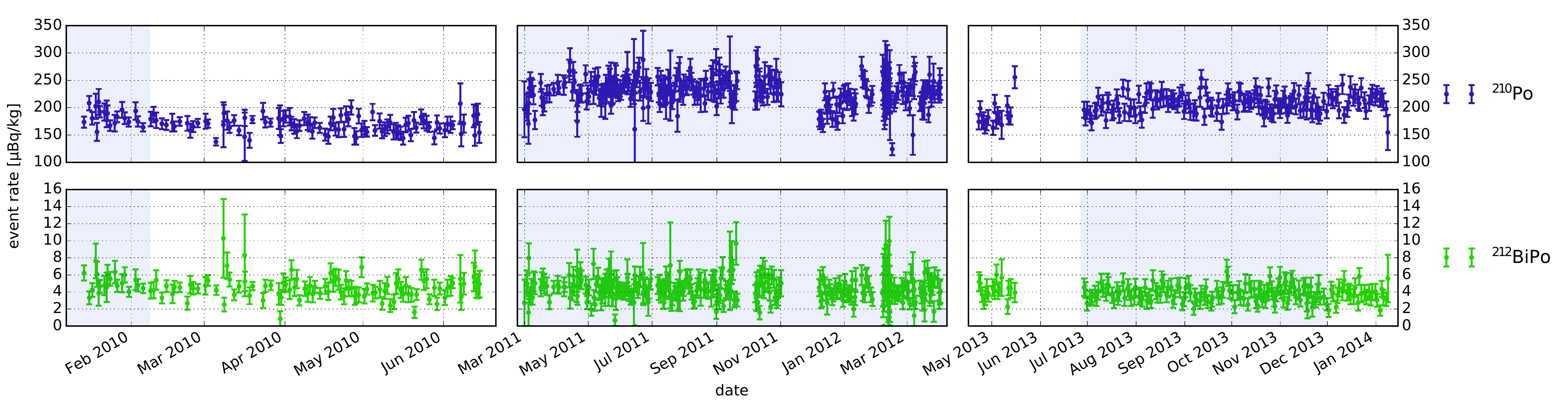}
	\caption{Radon event rates during the three SRs. Before starting SR1, \isotope[222]{Rn} entered the detector via an
	air leak which was subsequently closed. The decay of this additional radon contribution is clearly visible in the rates, and the decay constant
	is compatible with expectations from literature values (see text for more details). Note the rate correlation among
	\isotope[222]{Rn}, \isotope[218]{Po} (selected by $\alpha$-counting) and \isotope[214]{BiPos} (selected by delayed coincidence
	analysis) which are shown in the top row. The rates of \isotope[212]{BiPos} (delayed coincidence analysis) and \isotope[210]{Po} (wall event
	selection), shown in the bottom rows, do not correlate with them. Regions affected by a leak are shaded in blue.
	Mismatches between rates of the \isotope[222]{Rn} chain nuclides are explained in the text}
	\label{fig:rn_rates}
\end{figure*}

The rate evolution of each decay is shown in Figure~\ref{fig:rn_rates}. To verify that the selected event populations represent the
correct nuclei, an exponential decay plus a constant offset is fitted to the rate decrease of \isotope[222]{Rn} that is observed
during the two months of SR1 (Figure~\ref{fig:rn_rates}, top left). A small air leak was closed before this period, resulting in the decay
of the excess radon which is visible in the rate evolution.

The half-life given by the fit is $T_{1/2} = \SI{3.81(12)}{\day}$, which is in perfect agreement with the
literature value for \isotope[222]{Rn} ($T_{1/2} = \SI{3.82}{\day}$). In addition, the relative positions of the peaks in the
S1$_{\alpha}$ spectrum match the expectations given by the $\mathcal{Q}$-values of the individual decays,
with a constant light yield of $\sim \SI{3.7}{\photoelectron\per\kilo\electronvolt}$ observed for all nuclides. Furthermore, the rates assigned
to \isotope[222]{Rn}, \isotope[218]{Po} and \isotope[214]{BiPo} (radon chain) correlate with each other, while no correlation with
the rates from \isotope[212]{BiPo} (thoron chain) and \isotope[210]{Po} (radon chain) can be seen. While the latter is also part of the radon chain, secular equilibrium is broken
due to the long half-life of \isotope[210]{Pb} ($T_{1/2} = \SI{22.2}{y}$). While thoron can also enter the detector via leaks, it has a much shorter
half-life ($T_{1/2} = \SI{55.8}{\second}$) than \isotope[222]{Rn}, which results in a large suppression as it is more likely that it decays before reaching the TPC~\cite{Aprile:2016pmc}.

\begin{figure*}
	\centering
	\includegraphics[keepaspectratio=true, width=\textwidth]{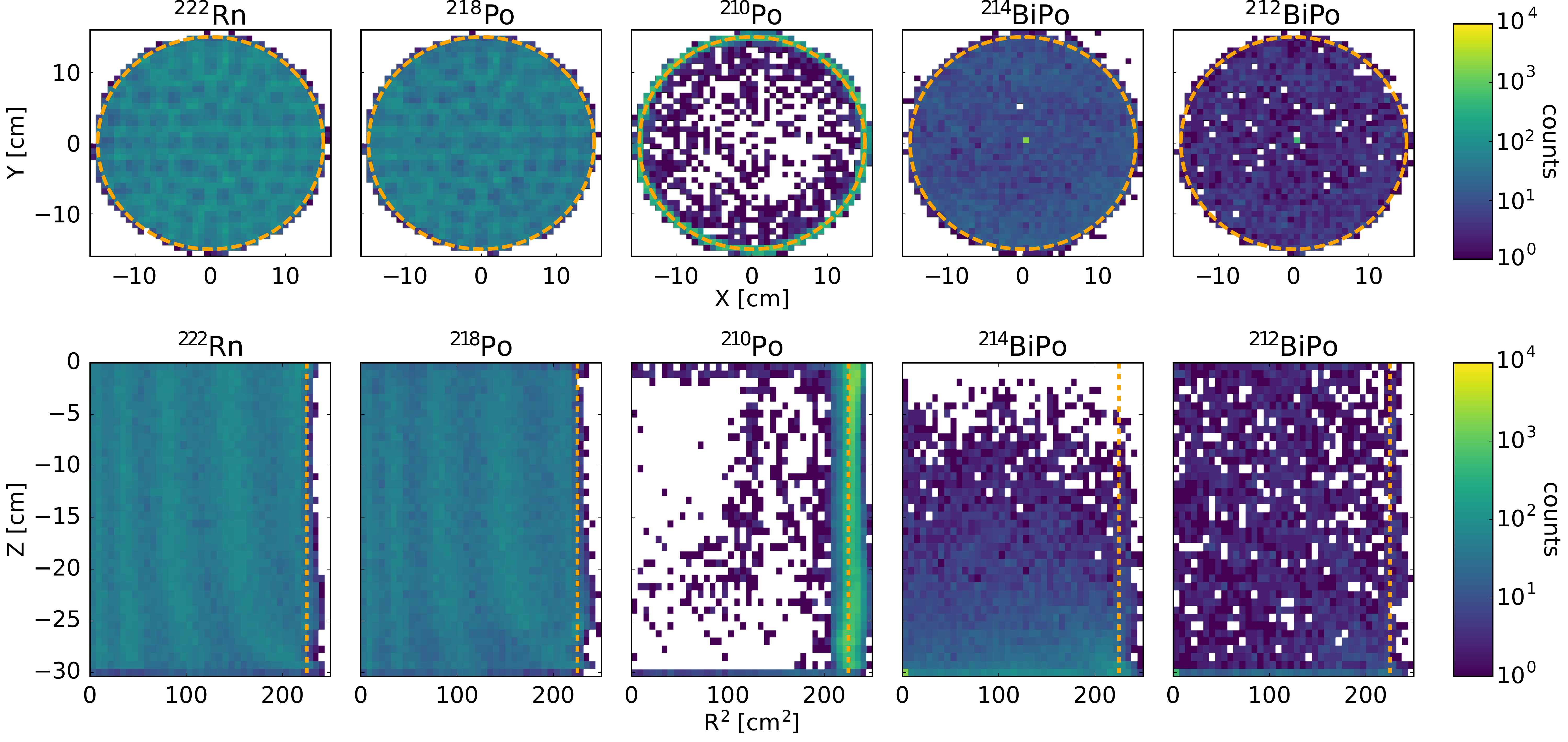}
	\caption{Spatial distributions of the radon populations identified in XENON100. Dashed lines indicate the TPC's radius. For
	\isotope[222]{Rn} and \isotope[218]{Po}, the main S2 signal is required to be larger than \SI{80000}{\photoelectron} to
	separate both from \isotope[210]{Po} at $R > \SI{135}{\milli\meter}$ (losses introduced by this cut are negligible at smaller radii).
	(Top) $XY$ distributions. Due to vertex reconstruction artifacts caused by PMT saturation, the structure of the top PMT array is
	visible in the \isotope[222]{Rn} and \isotope[218]{Po} data. (Bottom) $R^2Z$ distributions. Only events whose reconstructed
	$Z$ is within the TPC's actual height are shown. The same reconstruction artifacts as for the top plots are visible in the
	\isotope[222]{Rn} and \isotope[218]{Po} distributions. The number of \isotope[214]{BiPo} events seems to diminish toward the
	top of the TPC, but this is actually an artifact induced by the peak
	finder peculiarities described in \protect{Sec.~\ref{subsec:bipo_event_selection}} and by multi-scatters induced by $\gamma$-rays which accompany the
	\isotope[214]{Bi} decay\protect{~\cite{Weber:2013tci}}}
	\label{fig:spat_dists}
\end{figure*}

The condition on the S2 peak size introduced to select decays originating from the TPC's PTFE walls does not specifically select \isotope[210]{Po}.
However, considering that its rate is not correlated with the remainder of the radon chain, and taking into account similar
observations made by the LUX experiment~\cite{Bradley:2015ina}, we conclude that the wall population indeed consists of
\isotope[210]{Po}. The spatial distribution that includes it (see Figure~\ref{fig:spat_dists}) shows that it is located almost exclusively at $R^2 > \SI{200}{\centi\meter\squared}$, while the largest fiducial volume used for XENON100 WIMP analyses requires $R^2 < \SI{200}{\centi\meter\squared}$ among other constraints~\cite{Aprile:2011hi}.
For $R^2 < \SI{180}{\milli\meter\squared}$, a small number of events, likely caused by \isotope[222]{Rn} and \isotope[218]{Po} leakage,
can be seen. However, it is evident that these events are negligible compared to those happening at $R^2 \geq \SI{180}{\milli\meter\squared}$ as well as to those belonging to \isotope[222]{Rn} and \isotope[218]{Po}.

\sloppy
Computing the average specific rates (Table~\ref{tab:rates}) yields \SI{48.0(4)}{\micro\becquerel\per\kilo\gram},
\SI{64.3(4)}{\micro\becquerel\per\kilo\gram} and \SI{68.3(4)}{\micro\becquerel\per\kilo\gram} for \isotope[222]{Rn} in SR1 to SR3 respectively.
Periods of increased average rates and fluctuations are observed in SR2 and SR3. These increases are caused by tiny air leaks in the
diaphragm pump used in the xenon purification system, leading to a correlation of the \isotope[222]{Rn} rates inside and outside of the
detector~\cite{Aprile:2017yea}. Restricting the rate average to periods not affected by a leak gives
\SI{38.3(4)}{\micro\becquerel\per\kilo\gram} (SR1) and \SI{41.8(9)}{\micro\becquerel\per\kilo\gram} (SR3). We thus conclude that
constant emanation of radon from detector materials results in a base rate of, on average, \SI{40}{\micro\becquerel\per\kilo\gram}.

A direct measurement of the \isotope[222]{Rn} emanation at room temperature by means of miniaturized proportional counters was performed in summer 2012 between SR2 and SR3~\cite{Lindemann:2013}. It resulted in
\SI{9.3(10)}{\milli\becquerel} and \SI{2.6(5)}{\milli\becquerel} being measured for the XENON100 detector and gas system,
respectively, leading to an expected specific rate of \SI{74(7)}{\micro\becquerel\per\kilo\gram}
assuming homogeneous mixing of \isotope[222]{Rn} in the full LXe inventory. Inside the TPC, the assumption of homogeneous mixing is valid,
with the exception of \isotope[210]{Po} (Figure~\ref{fig:spat_dists}). The apparent decrease of \isotope[214]{BiPo} events
towards the top of the TPC is caused by losses induced by the peak finding algorithm as explained in
Sec.~\ref{subsec:bipo_event_selection} and by $\gamma$-rays, which accompany the \isotope[214]{Bi} decay, scattering off the
LXe at a different position than the original decay. Measurements with an external \isotope[222]{Rn} source suggest, that the homogeneous admixing of
radon throughout the entire LXe inventory takes place within a few hours~\cite{Aprile:2017kop}. The environmental conditions of the direct measurements
with proportional counters differed from the standard operation conditions, as the detector and gas system were at different temperatures and exposed
to nitrogen or helium, respectively. Both the increased temperature and reduced stopping power are known to impact the emanation rate of
\isotope[222]{Rn} (for example, see \cite{Sasaki:2004,Lindemann:2011aip}) and we consider the direct measurement to be a weak confirmation of our results.


\begin{table*}
	\centering
	\begin{tabular}{
	l
	S[table-format=3.2, table-number-alignment=right]@{\,\( \pm \)\,}
         S[table-format=2.2, table-number-alignment=left]
         S[table-format=3.2, table-number-alignment=right]@{\,\( \pm \)\,}
         S[table-format=2.2, table-number-alignment=left]
         S[table-format=3.2, table-number-alignment=right]@{\,\( \pm \)\,}
         S[table-format=2.2, table-number-alignment=left]
         S[table-format=3.2, table-number-alignment=right]@{\,\( \pm \)\,}
         S[table-format=2.2, table-number-alignment=left]
         S[table-format=3.2, table-number-alignment=right]@{\,\( \pm \)\,}
         S[table-format=2.2, table-number-alignment=left]
         S[table-format=3.2, table-number-alignment=right]@{\,\( \pm \)\,}
         S[table-format=2.2, table-number-alignment=left]}
	\toprule
	type&\multicolumn{12}{c}{rate [\si{\micro\becquerel\per\kilogram}]}\\
	&\multicolumn{2}{c}{SR1}&\multicolumn{2}{c}{SR1 (aft. leak)}&\multicolumn{2}{c}{SR2}&\multicolumn{2}{c}{SR3}&\multicolumn{2}{c}{SR3 (bef. leak)}&\multicolumn{2}{c}{SR3 (dur. leak)}\\
	\midrule
	\isotope[222]{Rn}&48.0&0.4&38.3&0.4&64.3&0.4&68.3&0.4&41.8&0.9&76.7&0.4\\
	\isotope[218]{Po}&41.0&0.4&33.4&0.4&52.2&0.3&59.0&0.3&37.1&0.9&66.1&0.4\\
	\isotope[210]{Po}&171.0&1.4&168.4&1.5&229.9&1.3&205.6&1.2&185&4&206.7&1.4\\
	\isotope[214]{BiPo}&24.8&0.6&20.5&0.6&32.8&0.5&36.9&0.5&22.9&1.2&41.1&0.6\\
	\isotope[212]{BiPo}&4.59&0.11&4.48&0.12&4.41&0.08&3.88&0.08&4.3&0.3&3.86&0.09\\
	\bottomrule
	\end{tabular}
	\caption{Average specific rates for all SRs (statistical errors only). The leak period of SR1 ends on February 7, 2010, and the leak
	period of SR3 lasts from June 27, 2013, to December 1, 2013. Note that the \isotope[210]{Po} rate concentration is large compared
	to the other nuclides because it is concentrated at the PTFE wall enclosing the TPC}
	\label{tab:rates}
\end{table*}

\textit{A priori}, we expect the radon chain from \isotope[222]{Rn} to \isotope[214]{Po} to be in secular equilibrium, as the longest-lived
daughter nuclide in this part of the chain, \isotope[214]{Pb}, has a half-life of \SI{26.9}{\minute} (Figure~\ref{fig:rn_chains}). This is
short compared to both the time scales of the SRs, which lasted for several months (Figure~\ref{fig:rn_rates}), and the time scale of the target purification, which is about \num{5} days per revolution. However, we observe only
about \SI{50}{\percent} of the expected amount of \isotope[214]{BiPo} events and about \SI{86}{\percent} of \isotope[218]{Po} events (Table~\ref{tab:rates}).
Acceptance losses due to cuts are negligible for the $\alpha$-events from \isotope[218]{Po} because of their high-energy signature. Thus, we have to consider additional causes for this mismatch. The most appealing one is radon daughters plating out onto
the cathode due to convection and drift in the electric field. Radon daughters which remain ionized were, for example, observed in the
EXO-200 TPC~\cite{Albert:2015vma}, and plating of radon progeny onto the cathode of a LXe TPC has already been reported by the ZEPLIN-III collaboration~\cite{Araujo:2011as}. The precise motivation for the plate-out hypothesis is the observation of a surplus of events in the
cathode region for \isotope[214]{BiPos} and \isotope[212]{BiPos} (Figure~\ref{fig:spat_dists}), which is visible even when rejecting events
without a proper S2 signal (which we assign to $Z = \SI{-30.5}{\centi\meter}$, the height of the cathode, by default). In addition,
it has been observed in \isotope[220]{Rn} calibration data, that the drift field affects the motion of \isotope[220]{Rn} daughters inside
the detector~\cite{Aprile:2016pmc}. While nuclide velocities inside the TPC are dominated by convection, which contributes up to $\sim~\SI{5}{\milli\meter\per\second}$ to up-/downward motion along the Z axis, a constant contribution of $\sim~\SI{1}{\milli\meter\per\second}$ towards the cathode is observed which is attributed to the drift field (\SIrange{500}{533}{\volt\per\centi\meter} depending on the SR~\cite{Aprile:2016swn}). As a consequence of the plate-out,
decays happening on the cathode are shadowed, leading to losses in the S1 signal and thus a lower acceptance of BiPo and \isotope[218]{Po} events.

No cathode accumulation is seen in the \isotope[218]{Po} distribution. Such an effect could be hidden due to the reduced discrimination
power at the bottom of the TPC between \isotope[218]{Po} and \isotope[222]{Rn} (see Sec.~\ref{subsec:alpha_event_selection}).
In addition, the effect on \isotope[214]{BiPos} is assumed to be enhanced because of the repeated chance of collecting ionized daughters
with every decay. A larger fraction of \isotope[214]{Bi} remaining ionized compared to \isotope[218]{Po}, as suggested in~\cite{Albert:2015vma}, might also play a role.

\isotope[210]{Po} rates are larger compared to those of other radon chain nuclides by a factor of $\sim 4.2$ in the outermost part of the detector
in periods not affected by a leak. However, one has to take into account that the volume within which \isotope[210]{Po} is selected is by a factor of $\sim 4.5$
smaller than a volume without any requirement on $R$ (Sec.~\ref{subsec:alpha_event_selection}). Averaging the \isotope[210]{Po} activity
without constraining $R$ gives, for instance, \SI{38.6(3)}{\micro\becquerel\per\kilo\gram} in SR1. Because this rate is still larger than the one observed for
\isotope[214]{BiPos} and does not correlate with rates of preceding chain decays, we assume surface contamination of the PTFE walls due to air exposure during TPC assembly to be the origin of the \isotope[210]{Po} population (analogous to observations made in~\cite{Clemenza:2011zz}). Under this assumption, we find a \isotope[210]{Po} activity per unit area of PTFE in the range from \SIrange{0.6}{0.9}{\micro\becquerel\per\cm\squared}.

\section{Krypton}
\label{sec:kr}
Natural krypton is present at the  parts-per-billion (ppb) level in
commercially available xenon produced in air separation plants.  It
contains the radioisotope \isotope[85]{Kr}, which is an almost pure
beta emitter and has a relatively long half-life of 10.76 years. Krypton
spreads throughout the liquid xenon target where it can induce
low-energy events that may leak into the WIMP search region. To mitigate the
\isotope[85]{Kr}-induced background, the
xenon target typically is purified by means of adsorption or distillation before starting a measurement~\cite{Bolozdynya:2007zz,Abe:2008py,Wang:2014ehv,Aprile:2016xhi}.  However,
re-contamination due to even tiny air leaks readily increases the
concentration of \isotope[85]{Kr}.

Natural sources result in a constant equilibrium content of 0.09 PBq
\isotope[85]{Kr} in the atmosphere~\cite{Schroeder1975}. In addition,
\isotope[85]{Kr} is produced alongside plutonium in spent nuclear
fuel and irradiated breeding targets. The noble gas remains therein until
it is released by nuclear fuel reprocessing facilities during the
extraction of plutonium. These anthropogenic sources increase the
atmospheric concentration of \isotope[85]{Kr} by orders of magnitude~\cite{Winger2005183}. In
present-day northern atmosphere, the activity of \isotope[85]{Kr} is
approximately \SI{1.4}{\becquerel\per\cubic\meter}~\cite{Loosli:2000,Bieringer:2009}.
This number roughly corresponds to a
relative isotopic abundance of \isotope[85]{Kr}/\isotope[\text{nat}]{Kr} =
\SI{2e-11}{\mole\per\mole}~\cite{Du:2003}.
However, the \isotope[85]{Kr} concentration varies across both time and space due to location and duty cycles of reprocessing plants as well as region-specific meteorological conditions~\cite{Winger2005183}.
\sloppy
To our knowledge, no atmospheric \isotope[85]{Kr} monitoring data is
publicly available for the region around LNGS and for the relevant period
of time. A single measurement (October 1, 2009) using miniaturized proportional counters of an air sample drawn
underground close to the XENON100 detector exists, resulting in \SI{1.33
\pm 0.16}{\becquerel\per\cubic\meter}~\cite{Lindemann:2009diplm}, or
\isotope[85]{Kr}/\isotope[\text{nat}]{Kr} = \SI{2.11 \pm 0.25
e-11}{\mole\per\mole} in agreement with the expected average value. 

In the following section we will discuss an \textit{in situ} analysis technique to uniquely
identify \isotope[85]{Kr} decays, quantify the krypton abundance during the
investigated SRs, and compare the results to external measurements
using a gas chromatographic system and a rare gas mass spectrometer (RGMS)~\cite{Lindemann:2013kna}.

\subsection{Delayed coincidence analysis}
\label{subsec:kr_analysis}

\isotope[85]{Kr} disintegrates by $\beta^{-}$ emission to the
\isotope[85]{Rb} ground state or, in \perc{0.438} of all cases, to its
second excited level. The half-life of the latter is
\SI{1.015}{\micro\second}. This decay mode offers a unique feature for
\isotope[85]{Kr} identification. In  more than \perc{99} of all cases the
prompt $\beta^{-}$ emission with endpoint energy of \SI{173}{keV} is
followed by a single \SI{514}{keV} gamma. This clear, delayed coincidence
signature allows for an \textit{in situ} analysis of krypton concentrations in the
XENON100 detector despite the tiny branching ratio and low statistics.
Energy levels, branching ratios and half lives are taken from~\cite{Be:tab}.

A set of basic cuts is applied in order to reject electronic noise and to ensure data quality, closely following the procedure outlined in~\cite{Aprile:2012vw}. We require a twofold PMT coincidence level for both the largest and next-to-largest S1 signals, as well as a minimum width of both the S1 waveforms. In addition, no light must be seen by the PMTs observing the LXe volume outside of the TPC (veto volume) in coincidence with the two S1 signals. 
Background events due to increased electronic noise in SR2 and SR3 are also removed.
Finally, we require that at least one S2 signal be identified in each recorded event trace.

\isotope[85]{Kr} delayed coincidence events are selected by requiring that
the largest S1 (S1$_\gamma$) follows the next-to-largest S1
(S1$_\beta$) within a time window of \SIrange{0.5}{4.9}{\micro\second}. The
acceptance of this criterion is \perc{67.5}. In addition, we demand that the reconstructed S1$_\gamma$ and S1$_\beta$ energies
fall within amply defined energy ranges: for the gamma interaction this
is three times the detector resolution (taken from~\cite{Aprile:2011dd})
around the expected value, i.e., from \SIrange{330}{698}{keV}. The maximal
accepted S1$_\beta$ energy is \SI{219}{keV}, i.e., the decay's endpoint
energy of \SI{173}{keV} plus twice the detector's resolution.

Detector-specific acceptance losses for small energy S1$_\beta$ deposits
are avoided by requiring signals to exceed \SI{14}{\photoelectron},
corresponding to \SI{5.8}{keV}~\cite{Aprile:2015ibr}. The acceptance of the latter condition is computed
to be \SI{91.7}{\percent}, using the $\beta$-Fermi-Function and the GEANT4
implementation thereof~\cite{Venkataramaiah:1985,Agostinelli:2002hh,Hauf:2013}. Poisson-like
fluctuations in S1$_\beta$ that affect the transformation from energy to S1
are negligible compared to the remaining uncertainties and ignored in the
following. \isotope[212]{BiPo} events originating from the
\isotope[220]{Rn} decay chain (see Sec.~\ref{sec:rn}) close to or on the
PTFE wall that encloses the TPC constitute the background. These events
are successfully removed by requiring that the sum of all identified S2
signals, i.e., the ones from the $\beta$-particle and the
$\gamma$-particle, fall within the expected energy region of
\SIrange{514}{687}{keV}.  Conservatively, the region is enlarged by five
times the S2 energy resolution.

We use the energy and interaction-type dependent S1 light yield and S2 gain to
convert the energy ranges into S1 and S2 light signals. In SR1, the
statistics in \isotope[85]{Kr} events is sufficiently high in order to
determine both the S1 light yield at \SI{514}{keV_\gamma} and the S2 gain at
$\SI{514}{keV_\gamma} + \SI{48}{keV_\beta}$. The latter corresponds to the S2 sum signal of the
monoenergetic $\gamma$-particle and the $\beta$-electron with an average
energy of \SI{48}{keV}.
For our purpose, we can assume the S1 light yield and S2 gain are
constant throughout the three SRs investigated~\cite{Aprile:2016swn}.
Finally, we use NEST~\cite{Szydagis:2013sih}, evaluated at \SI{0.5}{kV\per\cm} similar to the XENON100 drift field, and the measured light yield at \SI{122}{keV},
to convert the upper bound of our acceptance window for $\beta$ particles into an S1 value.

To convert the number of identified delayed coincidence events into a krypton
concentration (always given in \si{\mole\per\mole}), we have to account for the lifetime, the
amount of xenon, the cut acceptances of \SI{61.7(20)}{\percent} in total, and the
relative isotopic abundance of \isotope[85]{Kr}. 
For simplicity, we assume \SI{2e-11}{\mole\per\mole} for the latter and resume the discussion
in the context of induced uncertainties at the end of this section.

\subsection{Results and discussion}
\label{subsec:kr_results}
\begin{figure*}
  \centering
  \includegraphics[keepaspectratio=true, width=0.95\textwidth]{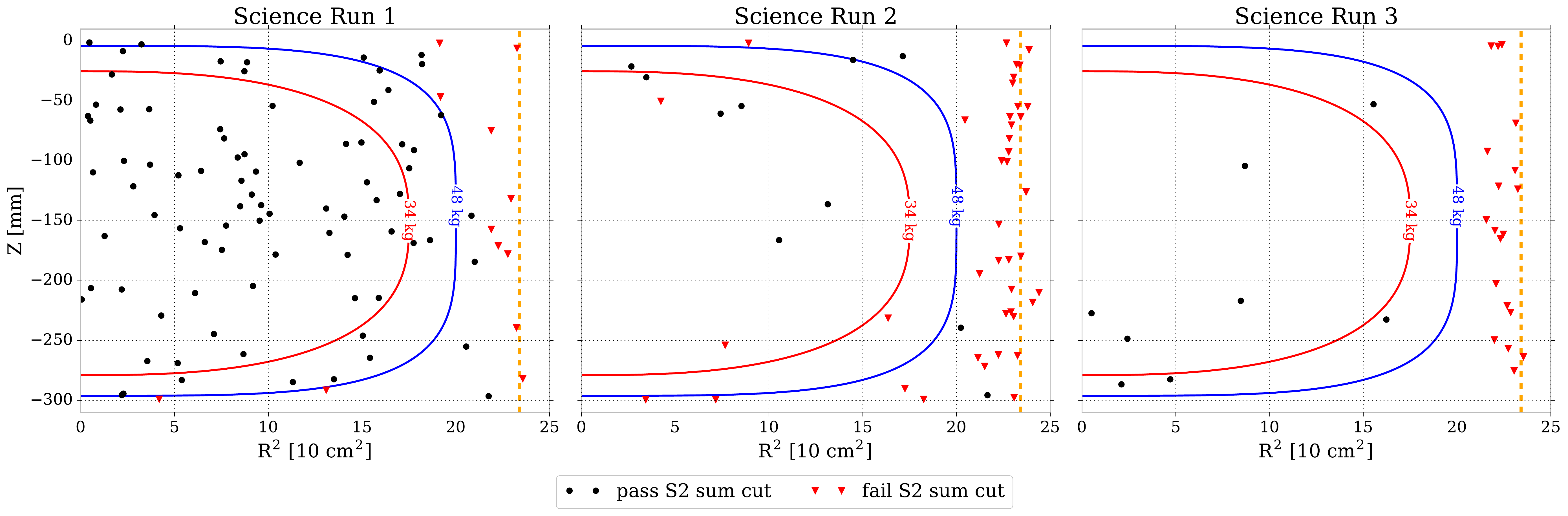}
  \caption{Observed event distributions for all three SRs.
    Displayed in black are events passing all cuts. Shown in red are those
    events that pass all selection cuts except for the condition on the S2 sum.
    See text for details. The red (34 kg) and blue (48 kg) contours indicate
    the two fiducial volumes used in the publications of the three SRs
	\protect{~\cite{Aprile:2011hi,Aprile:2012nq,Aprile:2016swn}}. Dashed lines indicate the TPC radius}
  \label{fig:event_distributions_rrz}
\end{figure*}

Figure \ref{fig:event_distributions_rrz} shows the event distributions
inside the TPC for the three SRs. Drawn in black are events passing all
data selection criteria. Plotted in red are the events that pass all
criteria except for the condition on the S2 sum. They are clearly clustered
close to the PTFE wall enclosing the TPC, while the former (black events)
are distributed throughout the TPC. For large radii, however, a reduced
acceptance for events passing all selection criteria becomes obvious. We
attribute this to a \SI{514}{keV} $\gamma$-ray's mean free path of roughly
\SI{2}{\centi\meter} in liquid xenon~\cite{nist:xraymasscoefz54}.  Close to
the wall these $\gamma$-rays can exit the TPC undetected and we lose the
characteristic pattern of the \isotope[85]{Kr} delayed coincidence. 
Fitting an exponential decay to the
time delay between S1$_\beta$ and S1$_\gamma$, we find $T_{1/2} = 1.08(_{-0.14}^{+0.18}) \,
\si{\micro\second}$ and $T_{1/2} = \SI{0.32 \pm 0.04}{\micro\second}$ for the events passing and
failing the S2 sum condition, respectively. This supports the hypothesis that the former (black events) indeed are due to \isotope[85]{Kr}, while the latter (red events) are
caused by the \isotope[212]{BiPo} delayed coincidence.

\begin{table}
  \centering
    \begin{tabular}{
            l
            l
            S[table-format=3.2, table-number-alignment=right]@{\,\( \pm \)\,}
            S[table-format=2.2, table-number-alignment=left]
        }
    \toprule
        period & date &  \multicolumn{2}{l}{\isotope[\text{nat}]{Kr}/Xe [ppt]} \\
    \midrule
            SR1 & 02 Jun 2010 & 340 & 60 \\
    \hline
            SR2 & 17 Nov 2011 & 13.8 & 2.4 \\
    \hline
         \multirow{4}{*}{SR3} & 14 Dec 2012 & 0.71 & 0.24 \\
		& 09 Jan 2013 & 0.95 & 0.22 \\
		& 21 Oct 2013 & 8.7 & 1.5 \\
		& 22 Dec 2013 & 11.1 & 1.9 \\
    \bottomrule
  \end{tabular}
    \caption{Overview of the \isotope[\text{nat}]{Kr} concentration measurements that were performed during the three SRs using mass spectrometry}
  \label{tab:kr_rgms_results}
\end{table}
\renewcommand{\arraystretch}{1}

\renewcommand{\arraystretch}{1.1}
\begin{table}
  \centering
    \begin{tabular}{
            c
            c
            c
            S[table-number-alignment=right,
            table-format=1.0, 
            table-space-text-pre=2,
            table-space-text-post=22]
            c
            }
    \toprule
     &  &  & \multicolumn{2}{c}{\isotope[\text{nat}]{Kr}/Xe [ppt]} \\
    \cmidrule(l){4-5}
        period & FV [kg] & events [1] & \multicolumn{1}{c}{DC} & \multicolumn{1}{c}{RGMS} \\ \hline
    \multirow{3}{*}{SR1} & 34 & 54 &  370{$_{-50}^{+60}$}  & \multirow{3}{*}{\num{340 \pm 60}} \\
                         & 48 & 74 &  360{$_{-40}^{+50}$}  & \\
                         & 62 & 83 &  310{$_{-30}^{+40}$}  & \\ \hline
    \multirow{3}{*}{SR2} & 34 &  5 &   15{$_{-7}^{+10}$}   & \multirow{3}{*}{\num{11.0 \pm 1.7}} \\
                         & 48 &  7 &   15{$_{-6}^{+8}$}    & \\
                         & 62 & 10 &   17{$_{-5}^{+7}$}    & \\ \hline
    \multirow{3}{*}{SR3} & 34 &  4 &   18{$_{-9}^{+14}$}   & \multirow{3}{*}{\num{6.3 \pm 1.0}} \\
                         & 48 &  8 &   25{$_{-9}^{+13}$}   & \\
                         & 62 &  8 &   20{$_{-7}^{+10}$}   & \\
    \bottomrule
  \end{tabular}
  \caption{Result of the delayed coincidence study for the three SRs and
    considering three different fiducial volumes (FV). The number of
    tagged events is converted into a krypton concentration (DC). 
      The concentrations can be compared to the corresponding SR-averaged
      off-line measurements (RGMS). See text for details}
  \label{tab:kr85_dc_results}
\end{table}
\renewcommand{\arraystretch}{1}

Table~\ref{tab:kr_rgms_results} lists all \isotope[\text{nat}]{Kr}
measurements that were performed off-line with the RGMS setup using gas
samples drawn from the purification loop of XENON100. In this
loop, about \SI{5}{slpm} of xenon are continuously evaporated from the liquid
xenon phase. Due to this large mass flow, we assume the Kr concentrations
of these gaseous samples to represent the liquid xenon target. Employing
the model describing the time evolution of the krypton concentration from~\cite{Aprile:2017yea} and using the available RGMS measurements, we compute
the run-averaged krypton concentrations of \SI{340\pm60}{ppt},
\SI{11.0\pm1.7}{ppt} and \SI{6.3\pm1.0}{ppt} for SR1
to SR3, respectively. 
Along with the RGMS-derived concentrations, Table~\ref{tab:kr85_dc_results}
lists the number of identified delayed coincidence events
and the resulting krypton concentrations (\isotope[\text{nat}]{Kr}/Xe DC) found in
the three SRs in the full TPC and two smaller fiducial
volumes. As discussed above, the chance to miss a delayed coincidence event increases with
radius due to \SI{514}{keV} gammas escaping the TPC undetected. To account
for this, we consider the innermost fiducial volume
(\SI{34}{\kilo\gram}) only, where we find
$370_{-50}^{+60}$ \si{ppt} (SR1), $15_{-7}^{+10}$ \si{ppt} (SR2)
and $18_{-9}^{+14}$ \si{ppt} (SR3), in good
agreement with the average concentrations derived from the RGMS
measurements. 
Limited statistics prevails in the uncertainties of the delayed coincidence
method. For example, in SR2, we select only 5 events in \SI{7.6}{\tonne\day}
of exposure. Comparing the measurements of SR1 where statistics is most
favorable, we find that the gas
samples drawn from the liquid in fact represent the entire xenon target.
However, there is a small indication of higher concentrations from the delayed coincidence
analysis in SR2 and SR3, i.e., we compute probabilities of \num{0.058}
(\num{0.30}) for finding 4 (5) or more
events in SR3 (SR2) based on the corresponding
RGMS-estimated concentrations.
This hints at a background gaining more importance with
reduced krypton concentration and increased exposure times
by, for example, random coincidences due to altered noise conditions~\cite{Aprile:2016swn}. Alternatively, underestimating the abundance of
\isotope[85]{Kr} that entered the detector through the air leaks in SR2 and SR3
could cause the surplus in events observed in the delayed coincidence analysis. In contrast to SR1, where the krypton was
introduced in a short period of time for which we have a direct measurement of the
\isotope[85]{Kr}/\isotope[\text{nat}]{Kr} ratio, plumes arriving from the
two nearest reprocessing plants La Hague, France, and
Sellafield, England, could alter the abundance of \isotope[85]{Kr} for SR2
and SR3. We account for such an effect by averaging the \isotope[85]{Kr} activity
concentration in ambient air monitored by the German Federal Office for
Radiation Protection (BfS)~\cite{bfs:2014} at Mount Schauinsland close to
Freiburg, Germany, during the relevant periods of time. We find
correction factors of \SI[retain-explicit-plus]{+10}{\percent}, \SI[retain-explicit-plus]{+30}{\percent} and
\SI[retain-explicit-plus]{+50}{\percent} with respect to our initial assumption of
\isotope[85]{Kr}/\isotope[\text{nat}]{Kr} = \SI{2e-11}{\mole\per\mole} 
for SR1 to SR3, respectively. The resulting krypton
concentrations are $340_{-50}^{+60}$ \si{ppt}, $12_{-5}^{+8}$ \si{ppt} and
$12_{-6}^{+9}$ \si{ppt},
increasing the probabilities for the observed number of delayed
coincidences to 0.17 (0.50) in SR3 (SR2).  We assume this estimate to serve
as a conservative upper limit only. The distance from the dominant sources
La Hague and Sellafield to the monitoring station at Mount Schauinsland is
only half of the distance to the underground laboratories.  Increased
\isotope[85]{Kr} concentrations due to reprocessing cycles are supposed to
be reduced at LNGS. In fact, simulations suggest variations in central
Italy to be only on the order of \SI{0.5}{\becquerel\per\cubic\metre}~\cite{Ross:2010}. Yet, for future experiments a local \isotope[85]{Kr}
monitoring station is desirable to reduce this large systematic uncertainty. 

\section{Summary and conclusions}
\label{sec:sum}
In this work, we presented techniques for selecting decays of the radon
(\isotope[222]{Rn}) and thoron (\isotope[220]{Rn}) chains and those of
\isotope[85]{Kr}. These methods allow us to estimate the contributions of the
involved nuclei to the ER background and to study the distribution of
background sources within the LXe target.  Furthermore, they provide
complementary values to those gained via direct measurements of the
\isotope[222]{Rn} emanation rate and the concentration of natural krypton
in the xenon target. 

\begin{table}
	\begin{tabu}{X[0.5]X[0.5]X[c]X[c]X[c]}
	\toprule
	source&meas.&\multicolumn{3}{c}{induced ER rate [\si{\milli\dru}]}\\
	&&SR1&SR2&SR3\\
	\midrule
	\multirow{3}{*}{\isotope[222]{Rn}}&\isotope[222]{Rn}&\num{1.392(12)}&\num{1.865(12)}&\num{1.981(12)}\\
	&\isotope[218]{Po}&\num{1.189(12)}&\num{1.514(9)}&\num{1.711(9)}\\
	&\isotope[214]{BiPo}&\num{0.719(17)}&\num{0.951(15)}&\num{1.070(15)}\\
	\multirow{2}{*}{\isotope[85]{Kr}}&DC&$14^{+2}_{-2}$&$0.6^{+0.4}_{-0.3}$&$0.7^{+0.5}_{-0.4}$\\
	&RGMS&\num{13(2)}&\num{0.43(7)}&\num{0.25(4)}\\
	\bottomrule
	\end{tabu}
	\caption{Estimates for the average ER background induced by the \isotope[222]{Rn} chain and
	\isotope[85]{Kr} (below \SI{100}{\kilo\electronvolt}, before applying ER/NR discrimination). Values are
	inferred from different measurements. Only delayed coincidence values for the \SI{34}{\kilo\gram} fiducial volume are used, as they are affected the least by acceptance losses as outlined in \protect{Sec.~\ref{subsec:kr_results}}}
	\label{tab:er_rates}
\end{table}

\sloppy 
Background rates are given in units of differential rate (mDRU = \SI{1e-3}{events\per(\kilo\gram \ day \
\kilo\electronvolt)}). Monte Carlo studies~\cite{Aprile:2011vb},
combined with the assumption of \isotope[85]{Kr}/\isotope[\text{nat}]{Kr} =
\SI{2e-11}{\mole\per\mole}, yield conversion factors of \SI{0.029}{\milli\dru\per(\micro\becquerel\per\kilo\gram)} and
\SI{0.039}{\milli\dru\per\ppt}
to relate \isotope[222]{Rn} and \isotope[\text{nat}]{Kr} concentrations to
ER rates, respectively.
The \isotope[220]{Rn} chain has not been simulated due to \isotope[222]{Rn}
and its daughters being more abundant as observed in data.
The contribution of the \isotope[222]{Rn} chain can be estimated by using
the observed \isotope[222]{Rn} rates. In the XENON100 science runs covered
by this work, the xenon purity was affected by three air leaks of different
leak rates. Consequently, we divide the radon-induced ER background into a
constant pedestal driven by emanation and a variable offset due to
\isotope[222]{Rn} leaking into the detector. In most parts of SR1, we do not
observe a variable component due to external radon and infer the pedestal
ER rate to be \SI{1.4}{\milli\dru}. In SR2 (SR3) we find the variable offset
to account for \SI{0.5}{\milli\dru} (\SI{0.6}{\milli\dru}) on average. This corresponds to
\SIrange{35}{40}{\percent} of the total \isotope[222]{Rn}-induced ER
background.

However, as we explained in Sec.~\ref{subsec:rn_results}, this
overestimates the induced ER background because of plate-out effects. With
\isotope[214]{Pb} being the most relevant $\beta$-emitter and
ER background source of the chain~\cite{Aprile:2015uzo}, the actual ER
background is smaller, assuming that the discrepancies between
the decay rates arise mostly due to plate-out. We find the background index
reduced to \SI{86}{\percent} (\SI{50}{\percent}) if we take
\isotope[218]{Po} (\isotope[214]{BiPo}) rates to assess the effective activity
concentration. Table~\ref{tab:er_rates} lists the radon contribution to the ER background
using the different assumptions. 

ER background from \isotope[85]{Kr} can be estimated via both RGMS and delayed coincidence
measurements. While the RGMS measurements are more precise than the delayed coincidence
measurements, we have to account for systematic uncertainties in the
\isotope[85]{Kr}/\isotope[\text{nat}]{Kr} ratio. Delayed coincidence
measurements suffer from limited statistics, especially in SR2 and SR3, but
constitute a direct measurement of the \isotope[85]{Kr} concentration which
does not rely on any assumption for the krypton ratio.

Based on the analysis procedures detailed in this work, we can
quantify the amount of air that entered through the leaks by means of the
two tracers radon and krypton. In all cases, we find the
\isotope[85]{Kr}-based estimate to be a factor of approximately two
lower than the one from \isotope[222]{Rn}. From measurements with a
spiked \isotope[222]{Rn} source, we know that within only two hours radon
homogeneously admixes throughout the entire LXe inventory~\cite{Aprile:2017kop}. The agreement between \textit{in situ} delayed
coincidence and external RGMS measurements suggests that we do not miss a
significant fraction of krypton in the liquid xenon target. 
To resolve the apparent tension, we conclude that krypton is enriched in
the gaseous part of the detector beyond the expected value of \num{\sim 10}~\cite{Rosendahl:2015eor}.

The contribution of \isotope[85]{Kr} in SR1 is \SI{14\pm2}{\milli\dru} --
one order of magnitude larger than \isotope[222]{Rn}.
Krypton removal by cryogenic distillation results in \isotope[222]{Rn}
being dominant in SR2 and SR3. For instance in SR2, \isotope[222]{Rn} and
\isotope[85]{Kr} contribute \SI{29}{\percent} and \SI{11}{\percent} to the
total ER background, respectively. This emphasizes the necessity to understand and
control the target-intrinsic backgrounds radon and krypton and, in
particular, it outlines the importance of radon screening and
dedicated material selection campaigns complemented by online radon removal
techniques for current and future liquid noble gas experiments like nEXO~\cite{Albert:2017hjq}, DarkSide-20k~\cite{Aalseth:2017fik}, LZ~\cite{Mount:2017qzi}, XENON1T, XENONnT and DARWIN~\cite{Aalbers:2016jon}.

\section*{Acknowledgments}
We gratefully acknowledge support from the National Science Foundation, Swiss National Science Foundation, Deutsche Forschungsgemeinschaft, Max Planck Gesellschaft, German Ministry for Education and Research, Netherlands Organisation for Scientific Research, Weizmann Institute of Science, I-CORE, Initial Training Network Invisibles (Marie Curie Actions, PITNGA-2011-289442), Fundacao para a Ciencia e a Tecnologia, Region des Pays de la Loire, Knut and Alice Wallenberg Foundation, Kavli Foundation, and Istituto Nazionale di Fisica Nucleare. We are grateful to Laboratori Nazionali del Gran Sasso for hosting and supporting the XENON project.

\end{document}

%% file: affiliations.tex
\newcommand{\bologna}{Department of Physics and Astrophysics, University of Bologna and INFN-Bologna, 40126 Bologna, Italy}
\newcommand{\chicago}{Department of Physics \& Kavli Institute of Cosmological Physics, University of Chicago, Chicago, IL 60637, USA}
\newcommand{\coimbra}{LIBPhys, Department of Physics, University of Coimbra, 3004-516 Coimbra, Portugal}
\newcommand{\columbia}{Physics Department, Columbia University, New York, NY 10027, USA}
\newcommand{\freiburg}{Physikalisches Institut, Universit\"at Freiburg, 79104 Freiburg, Germany}
\newcommand{\lngs}{INFN-Laboratori Nazionali del Gran Sasso and Gran Sasso Science Institute, 67100 L'Aquila, Italy}
\newcommand{\mainz}{Institut f\"ur Physik \& Exzellenzcluster PRISMA, Johannes Gutenberg-Universit\"at Mainz, 55099 Mainz, Germany}
\newcommand{\heidelberg}{Max-Planck-Institut f\"ur Kernphysik, 69117 Heidelberg, Germany}
\newcommand{\munster}{Institut f\"ur Kernphysik, Westf\"alische Wilhelms-Universit\"at M\"unster, 48149 M\"unster, Germany}
\newcommand{\nikhef}{Nikhef and the University of Amsterdam, Science Park, 1098XG Amsterdam, Netherlands}
\newcommand{\nyuad}{New York University Abu Dhabi, Abu Dhabi, United Arab Emirates}
\newcommand{\purdue}{Department of Physics and Astronomy, Purdue University, West Lafayette, IN 47907, USA}
\newcommand{\rpi}{Department of Physics, Applied Physics and Astronomy, Rensselaer Polytechnic Institute, Troy, NY 12180, USA}
\newcommand{\rice}{Department of Physics and Astronomy, Rice University, Houston, TX 77005, USA}
\newcommand{\stockholm}{Oskar Klein Centre, Department of Physics, Stockholm University, AlbaNova, Stockholm SE-10691, Sweden}
\newcommand{\subatech}{SUBATECH, IMT Atlantique, CNRS/IN2P3, Universit\'e de Nantes, Nantes 44307, France}
\newcommand{\torino}{INFN-Torino and Osservatorio Astrofisico di Torino, 10125 Torino, Italy}
\newcommand{\ucla}{Physics \& Astronomy Department, University of California, Los Angeles, CA 90095, USA}
\newcommand{\ucsd}{Department of Physics, University of California, San Diego, CA 92093, USA}
\newcommand{\wis}{Department of Particle Physics and Astrophysics, Weizmann Institute of Science, Rehovot 7610001, Israel}
\newcommand{\zurich}{Physik-Institut, University of Zurich, 8057  Zurich, Switzerland}
\newcommand{\lpnhe}{LPNHE, Universit\'e Pierre et Marie Curie, Universit\'e Paris Diderot, CNRS/IN2P3, Paris 75252, France}

%% file: ib_authorlist_epjc.tex
\author{E.~Aprile\thanksref{columbia}
        \and
        J.~Aalbers\thanksref{nikhef}
        \and
        F.~Agostini\thanksref{lngs,bologna}
        \and
        M.~Alfonsi\thanksref{mainz}
        \and
        F.~D.~Amaro\thanksref{coimbra}
        \and
        M.~Anthony\thanksref{columbia}
        \and
        F.~Arneodo\thanksref{nyuad}
        \and
        P.~Barrow\thanksref{zurich}
        \and
        L.~Baudis\thanksref{zurich}
        \and
        B.~Bauermeister\thanksref{stockholm}
% ...
\\
        ~(XENON Collaboration)\thanksref{xenon}
        }
\thankstext{lindemann}{e-mail: sebastian.lindemann@physik.uni-freiburg.de}
\thankstext{cichon}{e-mail: dominick.cichon@mpi-hd.mpg.de}
\thankstext{bern}{Also at Albert Einstein Center for Fundamental
Physics, University of Bern, 3012 Bern, Switzerland}
\thankstext{conrad}{Wallenberg Academy Fellow}
\thankstext{lopes}{Also at Coimbra Engineering Institute, Coimbra,
Portugal}
\thankstext{xenon}{email: xenon@lngs.infn.it}
\institute{\columbia \label{columbia}
           \and
           \nikhef \label{nikhef}
           \and
           \lngs \label{lngs}
           \and
           \bologna \label{bologna}
           \and
           \mainz \label{mainz}
           \and
           \coimbra \label{coimbra}
           \and
           \nyuad \label{nyuad}
           \and
           \zurich \label{zurich}
           \and
           \stockholm \label{stockholm}
% ...
}